\documentclass[preprint,showpacs,preprintnumbers,superscriptaddress]{revtex4}
\usepackage{amsmath,color,graphicx}

\begin{document}
\bibliographystyle{plain}
\def\la{\langle}
\def\barr{\begin{array}}
\def\earr{\end{array}}
\def\la{\langle}
\def\ra{\rangle}
\def\eps{\varepsilon}
\def\l{\left}
\def\r{\right}
\def\mev{\mbox{ MeV}}
\def\gev{\mbox{ GeV}}
\def\msbar{\overline{\mbox{MS}}}
\def\ln#1{\mbox{log}{\l( #1 \r)}}
\def\op{{\mathcal{O}}}
\def\ep{\epsilon}
\def\twospace{\mbox{ }\mbox{ }}
\def\ChPT{$\chi$PT~}
\def\dn{\frac{d^{n}k}{(2\pi)^{n}}}
\def\div{\frac{1}{\hat{\epsilon}}}
\def\ttb{\overline{t}_2}
\def\qu{{\vec{q}_1}}
\def\qd{{\vec{q}_2}}
\newcommand{\tr}{{\rm tr}}
\newcommand{\Tr}{{\rm Tr}}
\newcommand{\vev}[1]{\left\langle #1 \right\rangle}
\newcommand{\VEV}[3]{\left\langle #1\left| #2 \right| #3\right\rangle}

\newcommand{\vn}{\vec{n}}
\newcommand{\vx}{\vec{x}}
\newcommand{\vy}{\vec{y}}
\newcommand{\vp}{\vec{p}}
\newcommand{\vP}{\vec{P}}
\newcommand{\bk}{\mathbf{k}}
\newcommand{\vk}{\vec{k}}
\newcommand{\vkperp}{\vec{k}_\bot}
\newcommand{\vq}{\vec{q}}
\newcommand{\vtheta}{\vec{\theta}} 

\newcommand{\bq}{{\mathbf{q}}}
\newcommand{\bP}{\mathbf{P}}
\newcommand{\bp}{\mathbf{p}}
\newcommand{\bn}{\mathbf{n}}
\newcommand{\bx}{\mathbf{x}}
\newcommand{\by}{\mathbf{y}}
\newcommand{\bz}{\mathbf{z}}

\newcommand{\lm}{{\ell m}}
\newcommand{\lmp}{{\ell' m'}}

\newcommand{\cH}{{\cal H}}
\newcommand{\ch}{{\cal H}}
\newcommand{\co}{{\cal O}}
\newcommand{\cO}{{\cal O}}
\newcommand{\ca}{{\cal A}}
\newcommand{\cZ}{{\cal Z}}
\newcommand{\cJ}{{\cal J}}
\newcommand{\cK}{{\cal K}}
\newcommand{\cP}{{\cal P}}
\newcommand{\cM}{{\cal M}}
\newcommand{\cn}{{\cal N}}
\newcommand{\cD}{{\cal D}}
\newcommand{\cU}{{\cal U}}
\newcommand{\cL}{{\cal L}}
\newcommand{\cI}{{\cal I}}
\newcommand{\cA}{{\cal A}}
\newcommand{\cG}{{\cal G}}
\newcommand{\cC}{{\cal C}}
\newcommand{\vac}{\text{vac}}


\title{ Next to Leading Order Chiral Perturbation theory of $K \pi \to \pi$ and $K\to\pi\pi$ amplitudes }
\author{Changhoan Kim}
\affiliation{
Department of Physics, Columbia University
New York, NY 10027, USA}
\affiliation{
School of Physics and Astronomy, University of Southampton, Highfield, Southampton, SO17 1BJ, UK}

\vspace{5cm}

\begin{abstract}
It is shown that the low energy coefficients of the next-to-leading order (NLO) chiral perturbation theory needed to determine $\Delta I=1/2$, $K\to\pi\pi$ decay amplitudes can be fixed by calculating $K\pi\to\pi$ amplitudes on lattice.
Unlike using NLO $K\to\pi\pi$ amplitudes proposed by Laiho and Soni,  
simulating $K\pi\to\pi$ transitions on lattice does not require evaluations
 of $s$-channel disconnected diagrams which have been an obstacle in practice.
\end{abstract}

\maketitle
\newpage

\section{Introduction}
Chiral perturbation theory($\chi$PT) has been a useful tool in understanding the physics of mesons.
In particular, it has been employed to extrapolate important quantities to the physical 
 pion mass from lattice calculations performed using somewhat heavier mesons.

Employing leading order chiral perturbation theory in the lattice QCD calculations of nonleptonic kaon decays was proposed by Bernard {\it et al} in Ref.\cite{Bernard:1985tm}.
In this proposal, the effective weak operators are rewritten in terms of
 meson fields. The coefficients of those operators(low energy coefficients) are determined through lattice simulations where lattice calculations of $K\to\vac$ and $K\to\pi$ can be used to determine the more difficult $K\to\pi\pi$ amplitudes.
This calculation has been done\cite{Blum:2001xb,Noaki:2001un} with quenched ensembles.
However, the results did not show a good agreement with experimental observation.

Since then computing power has drastically improved and we can afford to generate unquenched lattice ensembles.
The chiral perturbation theory treatment has also been extended with the next-to-leading order calculation done by Laiho and Soni\cite{Laiho:2002jq}.
However, at next leading order $K\to\pi\pi_{I=0}$ calculations in addition to $K\to \pi$ and $K\to\vac$ on lattice are needed to determine all $K\to\pi\pi$ amplitudes with physical kinematics.   

Unfortunately, there are significant difficulties in lattice calculations of $K\to\pi\pi_{I=0}$ transitions.
Those difficulties basically result from the existence of $s$-channel disconnected diagrams. 
The numerical evaluation of those diagrams turn out to be very hard because of the exponential decay of signal to noise ratio.
In order to avoid the difficulties, the author have proposed to use $K\pi\to\pi$ transitions in Ref.\cite{Kim:2007ri}. 
The most notable advantage of using $K\pi\to\pi$ amplitudes is the elimination of $s$-channel disconnected diagrams.
Since $K^0\pi^-$ state is the highest weight state of $I=3/2$ states,
 the $I_z$ cannot be zero after the interaction with weak operators whose $I_z=-1/2$.    
This non-zero $I_z$ guarantees the absence of $s$-channel disconnected diagrams.
 A more detailed discussion about the difficulties can be found in Ref.\cite{Kim:2007ri}.


In this paper, the detailed $\chi$PT formulas for $K\pi\to\pi$ processes are presented. 
In the next two sections, we fix notation
 by expressing the effective weak operators in QCD as elements 
 of definite irreducible representation of the $SU(3)_L \otimes SU(3)_R$ chiral symmetry group and their corresponding expressions in terms of meson fields follow.
Then, it will be explicitly shown that the LECs needed to determine the physical $K\to\pi\pi$ decay amplitude at NLO in $\chi PT$ can be determined
 by analyzing the chiral expansion of $K\pi\to\pi$ amplitudes at simple kinematic points.
Those effects of finite volume are discussed which could be an issue when two particle states with non-zero relative momentum are considered.  


\section{ Effective Weak Operators }
The operator product expansion(OPE) can be used to express $K\to\pi\pi$ decay amplitudes 
 in terms of matrix elements of the $\Delta S=1$ weak 
 effective Hamiltonian,
\begin{equation}\label{1}
    \langle \pi \pi |{\cal H}_{\Delta S=1}|K\rangle =
    \frac{G_{F}}{\sqrt{2}}V_{ud} V^{*}_{us} \sum c_{i}(\mu)
    \langle \pi \pi|Q_{i}(\mu)|K\rangle.
\end{equation}
\noindent 
In essence, the OPE separates two important physical scales :
  the $c_{i}(\mu)$s, called Wilson coefficients, which contain the short distance physics which can be calculated by QCD and electroweak perturbative techniques   
 and the matrix elements, $\langle \pi \pi|Q_{i}(\mu)|K\rangle$
 which are determined by the long distance physics for which nonperturbative methods are required. A thorough discussion can be found {\it e.g.} in Ref.\cite{Buras:1993dy}.

The $Q_i(\mu)$ are four quark operators of which there are 10,
\begin{equation}\label{2}
    Q_{1}=\overline{s}_{a} \gamma_{\mu} (1-\gamma^{5}) u_{a}
    \overline{u}_{b}\gamma^{\mu} (1-\gamma^{5}) d_{b},
\end{equation}
\begin{equation}
Q_{2}=\overline{s}_{a} \gamma_{\mu} (1-\gamma^{5}) u_{b}
    \overline{u}_{b}\gamma^{\mu} (1-\gamma^{5}) d_{a},
\end{equation}
\begin{equation}
Q_{3}=\overline{s}_{a} \gamma_{\mu} (1-\gamma^{5}) d_{a} \sum_{q}
    \overline{q}_{b}\gamma^{\mu} (1-\gamma^{5}) q_{b},
    \end{equation}
    \begin{equation}
Q_{4}=\overline{s}_{a} \gamma_{\mu} (1-\gamma^{5}) d_{b} \sum_{q}
    \overline{q}_{b}\gamma^{\mu} (1-\gamma^{5}) q_{a},
    \end{equation}
     \begin{equation}
Q_{5}=\overline{s}_{a} \gamma_{\mu} (1-\gamma^{5}) d_{a} \sum_{q}
    \overline{q}_{b}\gamma^{\mu} (1+\gamma^{5}) q_{b},
     \end{equation}
     \begin{equation}
Q_{6}=\overline{s}_{a} \gamma_{\mu} (1-\gamma^{5}) d_{b} \sum_{q}
    \overline{q}_{b}\gamma^{\mu} (1+\gamma^{5}) q_{a},
    \end{equation}
    \begin{equation}
Q_{7}=\frac{3}{2} \overline{s}_{a} \gamma_{\mu} (1-\gamma^{5})
d_{a}
     \sum_{q} e_{q}
    \overline{q}_{b}\gamma^{\mu} (1+\gamma^{5}) q_{b},
    \end{equation}
    \begin{equation}
Q_{8}=\frac{3}{2} \overline{s}_{a} \gamma_{\mu} (1-\gamma^{5})
d_{b}
     \sum_{q} e_{q}
    \overline{q}_{b}\gamma^{\mu} (1+\gamma^{5}) q_{a},
     \end{equation}
      \begin{equation}
Q_{9}=\frac{3}{2} \overline{s}_{a} \gamma_{\mu} (1-\gamma^{5})
d_{a}
     \sum_{q} e_{q}
    \overline{q}_{b}\gamma^{\mu} (1-\gamma^{5}) q_{b},
    \end{equation}
    \begin{equation}
Q_{10}=\frac{3}{2} \overline{s}_{a} \gamma_{\mu} (1-\gamma^{5})
d_{b}
     \sum_{q} e_{q}
    \overline{q}_{b}\gamma^{\mu} (1-\gamma^{5}) q_{a}.
\end{equation}
The $ Q_{1} $ and $ Q_{2} $ are called  
current-current weak operators. The operators $Q_{3}-Q_{6} $ 
arise from QCD penguin diagrams whereas $ Q_{7}-Q_{10} $ are 
from electroweak penguin diagrams.

In order to use chiral perturbation theory, 
 these operators must be written in terms of elements of irreducible representations of the chiral symmetry group.
The relevant four-quark operators can be arranged into irreducible representations of the $SU(3)_L\otimes SU(3)_R$ chiral group with definite isospin 
 as follows :  
\begin{eqnarray}
{\cal X}_{\bf 27,1}^{(3/2)}~&=& (\bar{s}d)_L
\left[(\bar{u}u)_L-(\bar{d}d)_L\right]+(\bar{s}u)_L(\bar{u}d)_L ,\\
{\cal X}_{\bf 27,1}^{(1/2)}~&=& (\bar{s}d)_L\left[
(\bar{u}u)_L+2(\bar{d}d)_L-3(\bar{s}s)_L\right]+(\bar{s}u)_L(\bar{u}d)_L ,\\
{\cal X}_{\bf 8,1}^{(1/2)}~&=& (\bar{s}d)_L(\bar{u}u)_L-
(\bar{s}u)_L(\bar{u}d)_L ,\\
\tilde{\cal X}_{\bf 8,1}^{(1/2)}~&=&(\bar{s}d)_L\left[(\bar{u}u)_L
+2(\bar{d}d)_L+2(\bar{s}s)_L \right]+(\bar{s}u)_L(\bar{u}d)_L ,\\
{\cal Y}_{\bf 8,1}^{(1/2)}~&=&(\bar{s}d)_L\left[
(\bar{u}u)_R+(\bar{d}d)_R+(\bar{s}s)_R\right], \\
{\cal Y}_{\bf 8,1}^{(1/2) c}&=& \left\{(\bar{s}d)_L\left[
(\bar{u}u)_R+(\bar{d}d)_R+(\bar{s}s)_R\right] \right\}^c ,\\
{\cal Y}_{\bf 8,8}^{(3/2)} ~&=&(\bar{s}d)_L\left[
(\bar{u}u)_R-(\bar{d}d)_R\right]+(\bar{s}u)_L(\bar{u}d)_R, \\ 
{\cal Y}_{\bf 8,8}^{(3/2) c}&=& \left\{ (\bar{s}d)_L\left[
(\bar{u}u)_R-(\bar{d}d)_R\right] \right\}^c + 
 \left\{ (\bar{s}u)_L(\bar{u}d)_R \right\}^c, \\
{\cal Y}_{\bf 8,8}^{(1/2)}~&=&(\bar{s}d)_L\left[
(\bar{u}u)_R-(\bar{s}s)_R\right]-(\bar{s}u)_L(\bar{u}d)_R, \\ 
{\cal Y}_{\bf 8,8}^{(1/2) c} &=& \left\{ (\bar{s}d)_L\left[
(\bar{u}u)_R-(\bar{s}s)_R\right] \right\}^c
 - \left\{ (\bar{s}u)_L(\bar{u}d)_R \right\}^c.
\end{eqnarray}
\noindent
where we follow the notation of Ref.\cite{Noaki:2001un}.
Operators are classified by the Lorentz structure $L\otimes L$ and $L\otimes R$
 and represented by ${\cal X}$ and ${\cal Y}$ respectively.
The irreducible representation to which the operator belongs is given in the subscript and the short hand notation 
\begin{equation} 
(\bar{s}d)_L = \bar{s}^a \gamma_\mu(1 - \gamma_5)d^a\,, ~~~~
(\bar{s}d)_R = \bar{s}^b \gamma_\mu(1 + \gamma_5)d^b
\end{equation}
is used.
Because of invariance under the Fierz transformation,
 there are no color mixed $L\otimes L$ operator appearing,
 but for $L\otimes R$ operators, their color mixed versions are represented by superscript $c$ :
\begin{equation}
\left\{ (\bar{s}d)_L (\bar{s}d)_R \right\}^c = \bar{s}^a \gamma_\mu(1 - \gamma_5)d^b\,\bar{s}^b \gamma_\mu(1 + \gamma_5)d^a.
\end{equation}
Finally, the isospin of the operators is also given in the superscript.

In terms of this basis, the four-quark operators are rewritten as 
\begin{eqnarray}
Q_1&=&\frac{1}{2}{\cal X}_{\bf 8,1}^{(1/2)}+\frac{1}{10}
\tilde{\cal X}_{\bf 8,1}^{(1/2)}+\frac{1}{15}{\cal X}_{\bf 27,1}^{(1/2)}+
\frac{1}{3}{\cal X}_{\bf 27,1}^{(3/2)} \label{eq:q1} ,\\
Q_2&=&-\frac{1}{2}{\cal X}_{\bf 8,1}^{(1/2)}+\frac{1}{10}
\tilde{\cal X}_{\bf 8,1}^{(1/2)}+\frac{1}{15}{\cal X}_{\bf 27,1}^{(1/2)}+
\frac{1}{3}{\cal X}_{\bf 27,1}^{(3/2)} ,\\
Q_3&=&\frac{1}{2}{\cal X}_{\bf 8,1}^{(1/2)}+\frac{1}{2}
\tilde{\cal X}_{\bf 8,1}^{(1/2)} ,\\
Q_4&=& Q_2 + Q_3 - Q_1 ,\\
Q_5&=&{\cal Y}_{\bf 8,1}^{(1/2)} ,\\ 
Q_6&=&{\cal Y}_{\bf 8,1}^{(1/2)\ c} ,\\
Q_7&=&\frac{1}{2}\left[{\cal Y}_{\bf 8,8}^{(1/2)}+
{\cal Y}_{\bf 8,8}^{(3/2)}\right] ,\\
Q_8&=&\frac{1}{2}\left[{\cal Y}_{\bf 8,8}^{(1/2)\ c}+
{\cal Y}_{\bf 8,8}^{(3/2)\ c}\right] ,\\
Q_9&=& \frac{3}{2} Q_1 - \frac{1}{2} Q_3  ,\\
Q_{10}&=& Q_2 - \frac{1}{2} Q_3 + \frac{1}{2} Q_1 \label{eq:q10} .
\end{eqnarray}

The strategy is to measure the matrix elements of the $\cal X$ and $\cal Y$ operators with quarks somewhat heavier than physical quarks and 
 to use the chiral expansion to extrapolate the matrix elements to the physical quark mass. 
Then, we can recover the matrix elements of $Q_i(\mu)$s from Eqs.(\ref{eq:q1}-\ref{eq:q10}).


\section{Chiral Perturbation Theory}\label{sec:chiPT}
	Chiral perturbation theory ($\chi$PT) is based on  an effective field theory of the low energy sector of QCD.
Its fundamental degrees of freedom are the lowest mass pseudoscalar mesons which are the Goldstone-bosons arising from spontaneous chiral symmetry breaking. 
Because of the non-linear transformation of the Goldstone-bosons under the symmetry group, the meson fields appear in the field $\Sigma$ given by:
    \begin{equation}\label{3}
    \Sigma = \exp \left[\frac {2i\phi^{a} t^{a}}{f}\right],
\end{equation}
\noindent
 where $\Sigma$ belongs to the $(3,\bar{3})$ representation of the $SU(3)_L \otimes SU(3)_R$, the $3\times 3$ matrices $t^{a} $ are proportional to the Gell-Mann
matrices with $tr(t_a t_b)=\delta_{ab}$, and the $ \phi^{a} $ are the
real pseudoscalar-meson fields.
The quantity $f$ is the meson decay constant in the chiral limit, with $ f_{\pi} $
equal to 130 MeV in this notation \cite{Laiho:2002jq}.

The leading order ($O(p^{2})$) strong Lagrangian
is given by
\begin{equation}
   {\cal L}^{(2)}_{st}=\frac{f^{2}}{8}
   \textrm{tr}[\partial_{\mu}\Sigma^\dagger\partial^{\mu}\Sigma] +
   \frac{f^{2}B_{0}}{4}\textrm{tr}[\chi^{\dag}\Sigma+\Sigma^{\dag}\chi],
\end{equation}
\noindent where $\chi$ is a diagonal mass matrix with its diagonal elements $(m_u,m_d,m_s)$. 
They are related to the meson mass via
\begin{equation} 
    B_{0}= \frac{m^{2}_{\pi^{+}}}{m_{u}+m_{d}}=
    \frac{m^{2}_{K^{+}}}{m_{u}+m_{s}}=\frac{m^{2}_{K^{0}}}{m_{d}+m_{s}}.
\end{equation}
The leading order weak Lagrangian is given by
\begin{eqnarray}\label{5}
    {\cal L}^{(2)}_{W}&  =& 
    \alpha_{1}\textrm{tr}[\lambda_{6}\partial_{\mu}\Sigma^\dag\partial^{\mu}\Sigma]
    +\alpha_{2} 2B_{0}
    \textrm{tr}[\lambda_{6}(\chi^{\dag}\Sigma+\Sigma^{\dag}\chi)] \nonumber \\
    \!\!& & +
    \alpha_{27}t^{ij}_{kl}(\Sigma^\dag\partial_{\mu}\Sigma)^{k}_{i}
    (\Sigma^\dag\partial^{\mu}\Sigma)^{l}_{j} +
    \alpha_{88}\textrm{tr}[\lambda_{6}\Sigma^\dag Q \Sigma]
    + \textrm{H.c.}.
\end{eqnarray}
where terms with coefficients $\alpha_1$ and $\alpha_2$ belong to the $(8,1)$ representation of the $SU(3)_L \otimes SU(3)_R$ group and the last two terms belong to the $(27,1)$ and the $(8,8)$ representation respectively. 

The next-to-leading order($O(p^4)$) weak operators contributing
to kaon decays
 are given in
Ref.\cite{Golterman:2001yr}, with the effective Lagrangian, 
\begin{equation}\label{7}
    {\cal L}^{(4)}_{W}= \sum e_{i} {\cal O}^{(8,1)}_{i}+ \sum
    d_{i}{\cal O}^{(27,1)}_{i}
    +\sum c_{i} {\cal O}^{(8,8)}_{i}.
\end{equation}
The $(8,8)$ and $(27,1)$ operators will not be considered in this paper 
since the LECs for those operators can be determined without $\Delta I=1/2$, $K\to\pi\pi$ simulations \cite{Lin:2001ek}.  The explicit forms of the $ (8,1)$ operators are : 
\begin{equation}
\begin{array}{ll}
 {\cal O}^{(8,1)}_{1}= \textrm{tr}[\lambda_{6} S^2], & 
{\cal O}^{(8,1)}_{2}= \textrm{tr}[\lambda_{6} S] \textrm{tr}[S],
\\
{\cal O}^{(8,1)}_{3}=\textrm{tr}[\lambda_{6} P^{2}], &
{\cal O}^{(8,1)}_{4}=\textrm{tr}[\lambda_{6} P] \textrm{tr}[P],
\\
{\cal O}^{(8,1)}_{5}=\textrm{tr}[\lambda_{6}[S,P]], & 
{\cal O}^{(8,1)}_{10}=\textrm{tr}[\lambda_{6} \{S,L^{2}\}],
\\
{\cal O}^{(8,1)}_{11}=\textrm{tr}[\lambda_{6} L_{\mu} S L^{\mu}],
& 
{\cal O}^{(8,1)}_{12}=\textrm{tr}[\lambda_{6} L_{\mu}] \textrm{tr}[\{L^{\mu},S\}], \\
{\cal O}^{(8,1)}_{13}=\textrm{tr}[\lambda_{6} S] \textrm{tr}[L^{2}], &
{\cal O}^{(8,1)}_{15}=\textrm{tr}[\lambda_{6} [P,L^{2}]],\\
{\cal O}^{(8,1)}_{35}=\textrm{tr}[\lambda_{6}\{L_{\mu},\partial_{\nu}W^{\mu
\nu}\}], &
{\cal O}^{(8,1)}_{39}=\textrm{tr}[\lambda_{6} W_{\mu \nu} W^{\mu \nu}]
\end{array}
\nonumber
\end{equation}
\medskip
\noindent with
$ S=2B_{0}(\chi^{\dag}\Sigma + \Sigma^{\dag}\chi$),
$P=2B_{0}(\chi^{\dag}\Sigma-\Sigma^{\dag}\chi$),
$L_{\mu}=i \Sigma^{\dag}\partial_{\mu}\Sigma$ ,
$W^{\mu \nu}=2(\partial_{\mu}L_{\nu}+\partial_{\nu}L_{\mu})$,
 and $(\lambda_6)_{ij} = \delta_{3i} \delta_{2j}$.


Finally, the next-to-leading order strong Lagrangian relevant for kaon decay 
 amplitudes is
\begin{equation} 
\cL_{st}^{(4)} = \sum L_i \cO^{(st)}_i
\end{equation}
and the explicit forms for the $\cO^{(st)}_i$ are
\begin{equation} 
\begin{array}{ll}
{\cal O}^{(st)}_{1}=\textrm{tr}[L^{2}]^{2}, &
{\cal O}^{(st)}_{2}=\textrm{tr}[L_{\mu}L_{\nu}]\textrm{tr}[L^{\mu}L^{\nu}],\\
{\cal O}^{(st)}_{3}=\textrm{tr}[L^{2}L^{2}], &
{\cal O}^{(st)}_{4}=\textrm{tr}[L^{2}]\textrm{tr}[S],\\
{\cal O}^{(st)}_{5}=\textrm{tr}[L^{2}S],&
{\cal O}^{(st)}_{6}=\textrm{tr}[S]^{2},\\
{\cal O}^{(st)}_{8}=\frac{1}{2} \textrm{tr}[S^{2}+P^{2}]. &
\end{array}
\end{equation}

%
\section{Role of $K\pi\to\pi$ matrix elements}
%
In this section, we will show explicitly that those LECs necessary to calculate the physical $K\to\pi\pi$ amplitude can be determined from $K\to\vac$, $K\to\pi$ and $K\pi\to\pi$ amplitudes.

There are five types of operators depending on the representation of the chiral group and isospin : (27,1) $\Delta I=3/2$, (27,1) $\Delta I=1/2$,(8,8) $\Delta I=1/2$,(8,8) $\Delta I=3/2$ and (8,1) $\Delta I=1/2$.
In this paper, only (8,1) $\Delta I=1/2$ operators are discussed.
LECs for other operators (27,1)/(8,8) $\Delta I=1/2$ or $3/2$
 can be determined without the use of $K\to\pi\pi_{I=0}$ amplitudes\footnote{For a detailed explanation, see appendix \ref{apnd:C}}.

In order to check that all the LECs sufficient to reconstruct the physical $K\to\pi\pi$ amplitude can be determined, it is enough to look at the analytic terms.
Since matrix elements, meson masses and the meson momenta are calculated 
 on lattice, if we insert those quantities into the following $\chi$PT formulae a set of linear equations of LECs will be obtained. Presumably, the contribution of the logarithm terms would not render the linear equations singular.
So, we omit the logarithmic terms in the following formulae.

\begin{figure}[t]
\begin{center}
   \includegraphics[width=5in]{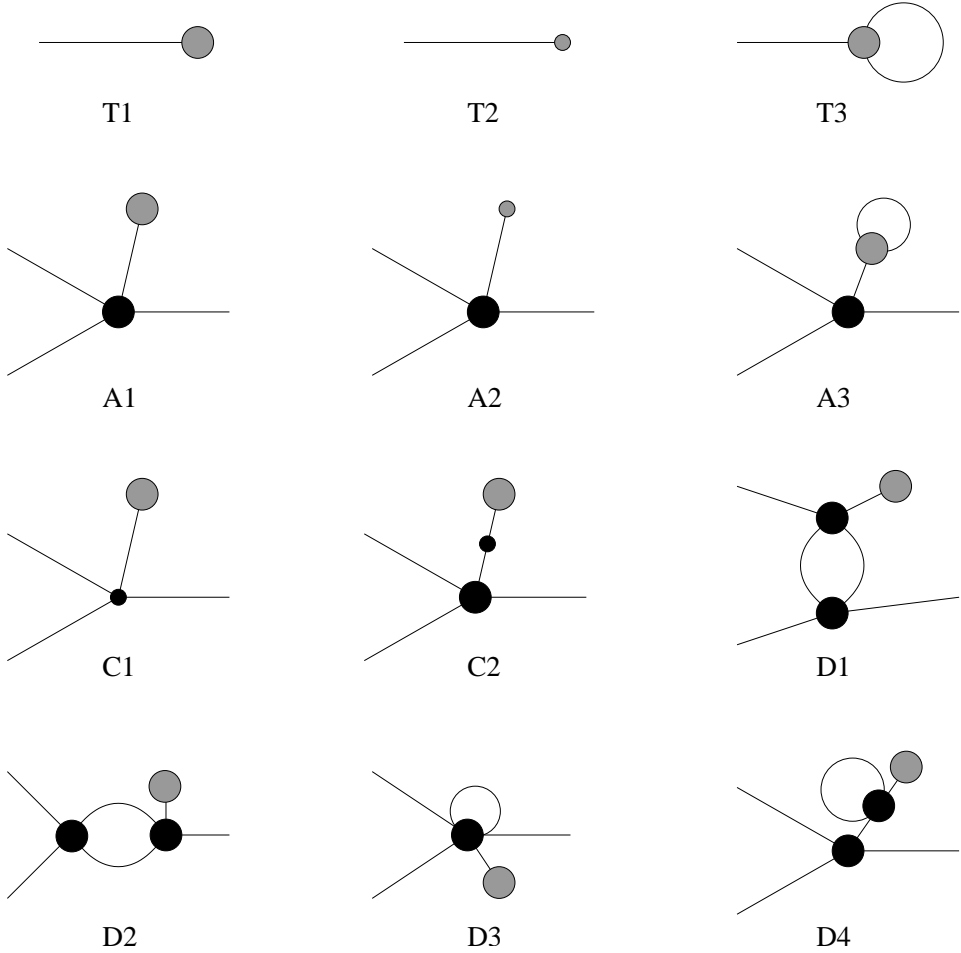}\\
\end{center}
\vspace{0.2cm}
\caption{Diagrams involved with tadpoles. T1,T2 and T3 are relevant to $K\to\vac$ amplitudes while others are for $K\pi \to \pi$ amplitudes. Gray(black) blobs are weak(strong) vertices. Large blobs are the leading order terms while the small blobs are the next-to-leading order ones.}
\label{fig:tad}
\end{figure}

Although $K\to\vac$ and $K\to\pi$ calculations have already been done, for example, in Ref.\cite{Laiho:2002jq},  we present the formulae for completeness.  
The analytic terms in the $K\to\vac$ amplitude come from diagrams T1 and T2 in Fig.\ref{fig:tad} and they are
\begin{equation} 
\vev{0|\cO^{(8,1)}|K^0}  =
-\frac{4 i \alpha _2 \left(m_K^2-m_{\pi }^2\right)}{f}
-\frac{8 i \left(m_K^2-m_{\pi }^2\right) \left(2 \left(e_1+e_2+e_5\right) m_K^2+e_2 m_{\pi
   }^2\right)}{f}.
\end{equation}
From this calculation, one can determine $\alpha_2$, $e_2$, $e_1 + e_5$.
If CPS symmetry\cite{Bernard:1985wf} is realized, $m_K = m_\pi$, the $K\to\vac$ matrix elements vanish. Thus, non-degenerate quark masses must be used for the determination of these LECs.

\begin{figure}[t]
\begin{center}
\includegraphics[width=5in]{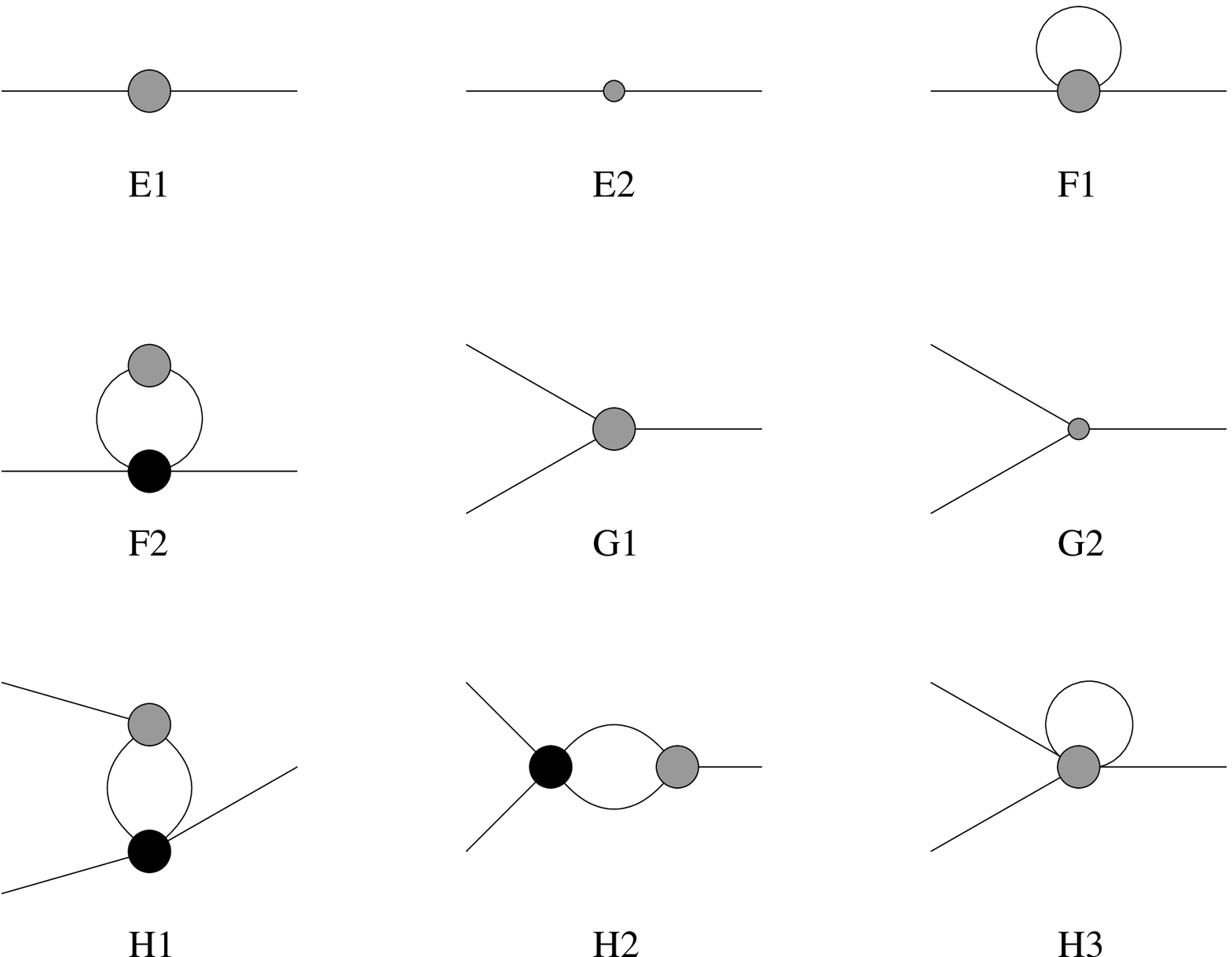} \\
\end{center}
\vspace{0.2cm}
\caption{Diagrams not involved with tadpoles. E1,E2 and F1 are relevant to $K\to\pi$ amplitudes while others are for $K\pi \to \pi$ amplitudes. Gray(black) blobs are weak(strong) vertices. Large blobs are the leading order terms while the small blobs are the next-to-leading order ones.}
\label{fig:notad}
\end{figure}

The diagrams E1 and E2 in Fig.\ref{fig:notad} generate the analytic terms in the $K\to\pi$ amplitudes, which are
\begin{eqnarray} 
\lefteqn{\vev{\pi^+(k_\pi)|\cO^{(8,1)}|K^+(p_K)} =
\frac{4 \alpha _1 (p_K \cdot k_\pi)-4 \alpha _2 m_K^2}{f^2} }
\nonumber \\&&
+\frac{1}{f^2} \Big(
-16 \left(e_1+e_2+e_5\right) m_K^4
\nonumber \\&&
+\left(16 \left(e_{10}-e_{35}\right) (p_K \cdot k_\pi)-8 \left(e_2+2
   e_3-2 e_5\right) m_{\pi }^2\right) m_K^2
\nonumber \\&&
+64 e_{39} (p_K \cdot k_\pi)^2
\nonumber \\&&
+8 \left(e_{11}-2 e_{35}\right) (p_K \cdot k_\pi)m_{\pi }^2
\Big) .
\end{eqnarray}
By varying momenta and masses, one can determine
$\alpha_1$, $e_{39}$, $e_{11} - 2 e_{35}$, $e_{10} -  e_{35}$
 and $e_3 - e_5$ when combined with $K\to\vac$ results. 

For $K\pi\to\pi$, there are five diagrams: A1, A2 and C1 in Fig.\ref{fig:tad} 
 and G1 and G2 in Fig.\ref{fig:notad}. The analytic terms are
\begin{eqnarray} 
\lefteqn{\vev{\pi^-(k_\pi) | \cO^{(8,1)}| K^0(p_K) \pi^-(p_\pi)}_{G1+G2} = \frac{i}{f^3} \Big( } && \nonumber \\
&& 
\frac{8}{3}  \alpha _2 \left(m_K^2-m_{\pi }^2\right)-4  \alpha _1 ((p_K \cdot k_\pi)-(p_\pi \cdot k_\pi))
\nonumber \\&&
+\frac{32}{3}  \left(e_1+e_2+e_5\right) m_K^4
+\frac{16}{3}  \left(e_1+5 e_2+3 e_3-2 e_5\right) m_{\pi }^2 m_K^2
\nonumber \\&&
-16  \left(\left(e_{10}-e_{35}\right) (p_K \cdot k_\pi)+\left(2 e_{13}+e_{15}\right) (p_\pi \cdot k_\pi)\right) m_K^2
\nonumber \\&&
-\frac{16}{3}  \left(3 e_1+7e_2+3 e_3\right) m_{\pi }^4
-64  e_{39} ((p_K \cdot k_\pi)^2-(p_\pi \cdot k_\pi)^2)
\nonumber \\&&
-8  \left(e_{11}-2\left(e_{15}+e_{35}\right)\right) (p_K \cdot k_\pi) m_{\pi }^2
\nonumber \\&&
+8  \left(2 e_{10}+e_{11}+4 e_{13}-4 e_{35}\right) (p_\pi \cdot k_\pi) m_{\pi}^2
\nonumber \\&&
+16  e_{35} (p_K \cdot p_\pi) ((p_K \cdot k_\pi)-(p_\pi \cdot k_\pi))
\Big),
\end{eqnarray}

\begin{eqnarray} 
\label{eq:A1A2}
\lefteqn{\vev{\pi^-(k_\pi) | \cO^{(8,1)}| K^0(p_K) \pi^-(p_\pi)}_{A1+A2} = } &&
\nonumber \\&&
\frac{4 i \left(-m_{\pi }^2 +2 (p_K \cdot p_\pi)+(p_K \cdot k_\pi)+(p_\pi \cdot k_\pi)\right) \left(m_{\pi }^2-m_K^2\right) \left( \alpha_2 + 4
   \left(  e_1+e_2+e_5\right) m_K^2+2e_2 m_{\pi }^2\right)}{3 f^3 \left(m_{\pi }^2+(p_K \cdot p_\pi)-(p_K \cdot k_\pi)-(p_\pi \cdot k_\pi)\right)}, \nonumber
 \\
\end{eqnarray}

and
\begin{eqnarray*}
\lefteqn{ \vev{ \pi^-( k_{\pi} ) | \co^{(8,1)} | K^0(p_K) \pi^-(p_{\pi} )}_{C1} = \frac{i 16 \alpha_2(m_K^2 - m_\pi^2)}{3 f^5 (p_k + p_\pi - k_\pi)^2 - m_K^2 } \Big( }&& \\ 
&& -12 k_{\pi }\cdot p_K k_{\pi }\cdot p_{\pi } \left(2 L_1+L_2+L_3\right)\\
&& +6 p_K\cdot p_{\pi } \left(k_{\pi }\cdot p_{\pi } \left(4 L_1+2 L_2+L_3\right)+k_{\pi }\cdot p_K \left(4 L_2+L_3\right)\right)\\
&& +2 k_{\pi }\cdot p_{\pi } \left(12 L_1+3 L_3-8 L_4-L_5\right) m_K^2-4 p_K\cdot p_{\pi } \left(2 L_4+L_5\right) m_K^2\\
&& -\left(2 L_4+L_5-2 \left(2 L_6+L_8\right)\right) m_K^4-2 k_{\pi }\cdot p_K \left(2 L_4+L_5\right) m_K^2\\
&& +2 k_{\pi }\cdot p_K \left(6 L_2+3 L_3+5 L_4\right) m_{\pi }^2-2 k_{\pi }\cdot p_{\pi } L_4 m_{\pi }^2\\
&& -\left(11 L_4+4 L_5-6 \left(5 L_6+2 L_8\right)\right) m_K^2 m_{\pi }^2-2 p_K\cdot p_{\pi } \left(6 L_2+8 L_4+3 L_5\right) m_{\pi }^2\\
&& +\left(L_4-L_5+2 \left(L_6+L_8\right)\right) m_{\pi }^4\Big),
\end{eqnarray*}

\noindent
where the $L_i$ are Gasser-Leutwyler coefficients and it is assumed that those coefficients are already known.

Since the analytic terms of $K\pi\to\pi$ amplitudes are quite complex,
 we isolate the new coefficients which must be determined
 from $K\pi\to\pi$ amplitudes by inserting LECs that can be computed from $K\to\vac$ and $K\to\pi$ amplitudes.
One can easily see that only three coefficients remain to be determined 
 which are from the G2 contribution,
\begin{eqnarray} 
\lefteqn{\vev{\pi^-(k_\pi)|\cO^{(8,1)}|K^0(p_K) \pi^-(p_\pi)}_{G2,\text{part}} = \frac{i}{f^3} \Big( } &&
\nonumber \\&&
16  e_{35} (p_K \cdot p_\pi) ((p_K \cdot k_\pi)-(p_\pi \cdot k_\pi))
\nonumber \\&&
-32  e_{13} (p_\pi \cdot k_\pi) \left(m_K^2-m_{\pi }^2\right)
\nonumber \\&&
-16  e_{15} \left((p_\pi \cdot k_\pi) m_K^2-(p_K \cdot k_\pi) m_{\pi }^2\right) \Big).
\end{eqnarray}
In order to show explicitly that the needed coefficients can be determined,
 we choose kinematics where the initial kaon and pion are nearly at rest
 while the final pion has momentum $2\pi/L$.
With these kinematic points,
\begin{eqnarray}
\lefteqn{\vev{\pi^-(k_\pi)|\cO^{(8,1)}|K^0(p_K) \pi^-(p_\pi)}_{G2,\text{part}} = } && \nonumber
\\&&
\frac{16 i W \left( ( e_{35}-e_{15} ) ( W_\pi m_K^2 - W_K m_\pi^2) - e_{13} ( E_W + W_\pi ) ( m_K^2 - m_\pi^2) \right)}{f^3} \nonumber \\
\label{eq:G2part}
\end{eqnarray}
where $W$ is the energy of final pion and $W_K$ and $W_\pi$ are the energies 
 of initial kaon and pion respectively.
Although very small, we take into account the momentum of the initial particles.
When $K\pi\to\pi$ transitions are simulated on lattice,
 this small momentum originates from interactions between the kaon and the pion and we have no control over the direction of this momentum.
An average over solid angle must be taken.
More discussion of this point is given in the section \ref{sec:FVE}.
This effect is already included in the above formula.
Now, one can see that $e_{35}-e_{15}$ and $e_{13}$ can be determined by varying $m_K$ and $m_\pi$.
Then, combined with $K\to\vac$ and $K\to\pi$ results, one can determine
 the physical $K\to\pi\pi$ matrix element :
\begin{eqnarray*} 
\lefteqn{\vev{ \pi^-\pi^+ | \cO^{(8,1)} | K^0 }_{PHYS} = } &&
\\&&
\frac{8i}{f^3} 
 \left(m_K^2-m_{\pi }^2\right) \cdot
\\&&
 \Big(\left(e_{10}-2 e_{13}-e_{15}\right) m_K^2
\\&&
+\left(-2 e_1-4 e_2-2 e_3+2 e_{10}+e_{11}+4 e_{13}-4 e_{35}+8 e_{39}\right) m_{\pi }^2\Big).
\end{eqnarray*}


Note that the SU(3) limit cannot be used because Eq.(\ref{eq:G2part}) vanishes in that limit.
Moreover, there is a further restriction on the choice of momenta if one considers more general kinematics.
The tadpole contribution(Eq.(\ref{eq:A1A2})) can diverge for some choices of momenta.
This is a disadvantage of using transitions with an unavoidable energy-momentum injection such as the current proposal.

%


\section{Finite Volume Effects}\label{sec:FVE}
In the previous section, it is shown that 
 the unknown LECs can be determined by measuring
 weak matrix elements from lattice calculations with mesons nearly at rest.
We choose to use a minimal set of matrix elements for the sake of proving sufficiency of the calculation of only these types of amplitudes.
However, it would be useful to measure more matrix elements rather than 
 the minimal set so that one can reduce statistical errors in determining the LECs and test for the consistency of the $\chi PT$ expansions.

An obvious approach is to include matrix elements with mesons of different masses,
 but this requires new sets of configuration if we want to use unquenched chiral perturbation theory.
A method which does not require new ensembles is to use mesons with non-zero relative momenta.
In this case, the finite volume effects on the resulting matrix elements are not exponentially small. This has been studied by L\"uscher and Lellouch in Ref.\cite{Lellouch:2000pv}
 and it is generalized to the case where the total momentum is not zero in Refs.\cite{Kim:2005gf,Christ:2005gi}.
This generalization is quite useful
 since using a system with non-zero total momentum is the only method, at the moment, of creating two particle states with non-zero relative momentum without generating new gauge ensembles. 

One can also consider matrix elements whose final meson states have non-zero momentum. 
In such cases, the result of Refs.\cite{Kim:2005gf,Christ:2005gi} has to be generalized further. 
This is straightforward and the derivation is given in appendix \ref{sec:apnd_FVE}. The final result is
\begin{eqnarray}
\lefteqn{ \Big|\vev{\pi(\bk) | H_W(0)  | K\pi,(E,\bP) }\Big| = }&&  \nonumber \\
&& \frac{1}{4\pi} \frac{1}{\sqrt{\rho_V}}
\sqrt{ \frac{\pi}{(2\pi)^3 } \left(\frac{q^*}{E^*} \right)} 
\Big|\int d\Omega_{E^*}
 \vev{  \pi(\bk^*)| H_W(0) |K(\bq) \pi(-\bq) }\Big|,
\label{eq:FVE_final}
\end{eqnarray}
where the left-hand side represents a matrix element in a finite box while the integrand on the right-hand side represents one in an infinite volume.  
Here, $E$ is the energy of the $K\pi$ state which can be measured from lattice calculations.
Similarly, $\bP$ is the total momentum of the $K\pi$ state which is imposed explicitly by the operator creating the state.
Finally, $\bk$ is the momentum of the final pion.
The starred variables are those Lorentz-transformed into the CM frame.
 In particular,     
\begin{equation} 
E^* \equiv \sqrt{E^2 - P^2}
\end{equation}
which in turn gives the Lorentz transformation angle $\beta=P/E$ and the $q^*$ is defined from  
\begin{equation}
 E^* =  \sqrt{ m_K^2 + q^{*2}} + \sqrt{ m_\pi^2 + q^{*2}}.
\end{equation}
The $\bk^*$ is the result of the Lorentz transformation with the above $\beta$ of the four-momentum of the final pion state, $(\sqrt{m_\pi^2 + \bk^2},\bk)$. 
$\rho_V$ is a function of $q^*$ and its definition is given in Refs.\cite{Kim:2005gf,Christ:2005gi}.
Roughly, it can be interpreted as the density of states.  

In Eq.(\ref{eq:FVE_final}), the difference from the result of Refs.\cite{Kim:2005gf,Christ:2005gi} is
 the explicit appearance of the integration over solid angle. Since the final meson state has non-zero momentum and cannot be used to constrain the initial state to be an S-wave, this must be done explicitly.
As such, the initial state in the matrix element $ \vev{\pi(\bk) | H_W(0)  | K\pi,(E,\bP) }$ must be in an S-wave state\footnote{see the appendix \ref{sec:apnd_FVE}}, as seen from the CM frame.
 This means that the S-wave state must be explicitly generated in the lattice simulation.
Fortunately, S-wave states are the lowest energy states which can appear, so we can generate the S-wave $K\pi$ state by using any operator which has an overlap with it and evaluating its correlation function at a large Euclidean time separation limit.

\section{Conclusion}
In this paper, it is shown that the LECs necessary for the NLO $\chi$PT calculation of the physical $\vev{\pi\pi|\cO^{(8,1)}| K }$ matrix elements
 can be obtained from lattice calculations of $K\pi\to\pi$,$K\to\pi$ and $K\to\vac$ processes. The important point is that simulations of $K\to\pi\pi_{I=0}$ transitions, which are very difficult, can be avoided. 
 
Although we establish this result by using a minimal set of kinematics,
 the finite volume effects on $K\to\pi\pi$ matrix elements are also discussed,
 effects which should be taken into account if one wants to improve the statistics and the control of systematic errors by including more kinematic points.
 In particular, the case where the final one particle state has non-zero momentum is discussed, which requires a slight generalization of the results of Ref.\cite{Kim:2005gf}.


As discussed in Ref.\cite{Kim:2007ri}, using $K\pi\to\pi$ processes allows us to avoid technical difficulties such as s-channel disconnected diagrams.
However, the mixing with lower dimensional operators is still present.
One may apply the subtraction scheme used in Ref.\cite{Blum:2001xb}.

The current computing resources are powerful enough to simulate $K\pi\to\pi$ transitions. With these NLO $\chi$PT formulae, we can improve the calculation of $K\to\pi\pi$ matrix elements, which in turn will allow us to evaluate $\epsilon^\prime / \epsilon$ more accurately.


\begin{acknowledgments}
The author thank Christopher Sachrajda and Norman Christ for useful discussions 
 and Micheal Endres and Matthew Lightman for reading the manuscript.
The author acknowledge that some of the calculations in this paper are done by \textsc{FeynArts} package \cite{Hahn:2000kx} on \textsc{Mathematica}.
This research is spported by PPARC 
grants PPA/G/O/2002/00468 and PPA/G/S/2003/00093, by DOE grant DE-FG02- 
96ER40956 and by the RIKEN-BNL Research Center. 
\end{acknowledgments}

\bibliography{myBibliography}

%
%
\appendix

\section{Logarithmic terms}
In this section, the logarithmic terms (diagram A3,F1,F2,D1,D2,D3,H1,H2,H3)
 are given.

\subsection{The Integrals}
In order to write the transition amplitude in a simpler form, the following notations are used :

\begin{eqnarray*} 
A_0(m) \equiv
  \mu^\epsilon \int \frac{d^d q }{ (2 \pi)^d } \frac{ 1 }{ q^2 - m^2 }  \\
A_1(m) \equiv
  \mu^\epsilon \int \frac{d^d q }{ (2 \pi)^d } \frac{ q^2 }{ q^2 - m^2 } 
\end{eqnarray*}

\begin{eqnarray*} 
B_0(m_1,m_2,k) \equiv
  \mu^\epsilon \int \frac{d^d q }{ (2 \pi)^d } \frac{ 1 }{ (q-k)^2 - m_1^2 } \frac{ 1 }{ (q+k)^2 - m_2^2 }  \\
B_{0\mu}(m_1,m_2,k) \equiv
  \mu^\epsilon \int \frac{d^d q }{ (2 \pi)^d } \frac{ q_\mu }{ (q-k)^2 - m_1^2 } \frac{ 1 }{ (q+k)^2 - m_2^2 }  \\
B_{0\mu\nu}(m_1,m_2,k) \equiv 
\mu^\epsilon \int \frac{d^d q }{ (2 \pi)^d } \frac{ q_\mu }{ (q-k)^2 - m_1^2 } \frac{ q_\nu }{ (q+k)^2 - m_2^2 }  \\
B_1(m_1,m_2,k) \equiv
 \mu^\epsilon \int \frac{d^d q }{ (2 \pi)^d } \frac{ q^2 }{ (q-k)^2 - m_1^2 } \frac{ 1 }{ (q+k)^2 - m_2^2 }  \\
B_{1\mu}(m_1,m_2,k) \equiv 
\mu^\epsilon \int \frac{d^d q }{ (2 \pi)^d } \frac{ q_\mu }{ (q-k)^2 - m_1^2 } \frac{ q^2 }{ (q+k)^2 - m_2^2 }  \\
B_2(m_1,m_2,k) \equiv 
\mu^\epsilon \int \frac{d^d q }{ (2 \pi)^d } \frac{ q^2 }{ (q-k)^2 - m_1^2 } \frac{ q^2 }{ (q+k)^2 - m_2^2 }  
\end{eqnarray*}

Furthermore, when there is a Lorentz contraction, we use the following abbreviation :
\begin{eqnarray*}
k \cdot B_0  \equiv B_{0\mu} k^\mu  \\
k \cdot B_1  \equiv B_{1\mu} k^\mu  \\
p \cdot B_0 \cdot k \equiv p^\mu B_{0\mu\nu} k^\nu 
\end{eqnarray*}

\subsection{Diagram H1a}
Here, $ q = \frac{1}{2}( p_\pi - k_\pi ) $, 
\begin{eqnarray*}
\lefteqn{ \vev{ \pi^-( k_{\pi} ) | \co^{(8,1)} | K^0(p_K) \pi^-(p_{\pi} )}_{H1a} = \frac{1}{f^5 \pi^2} \Big( }&& \\ 
&& 4 \alpha _1 k_{\pi }\cdot B_0\left(q,m_K,m_K\right)\cdot p_K+8 \alpha _1 k_{\pi }\cdot B_0\left(q,m_{\pi },m_{\pi }\right)\cdot p_K+4 \alpha _1 p_{\pi }\cdot B_0\left(q,m_K,m_K\right)\cdot p_K\\
&& +8 \alpha _1 p_{\pi }\cdot B_0\left(q,m_{\pi },m_{\pi }\right)\cdot p_K-\frac{2}{3} \alpha _2 \left(4 m_K^2-4 m_{\pi }^2\right) B_1\left(q,m_K,m_K\right)\\
&& +\frac{1}{3} \alpha _1 \left(-2 m_K^2-5 m_{\pi }^2+5 k_{\pi }\cdot p_{\pi }\right) \left(-m_{\pi }^2-k_{\pi }\cdot p_K+k_{\pi }\cdot p_{\pi }+p_K\cdot p_{\pi }\right) B_0\left(q,m_K,m_K\right)\\
&& +\frac{1}{3} \alpha _2 \left(-2 m_K^2-5 m_{\pi }^2+5 k_{\pi }\cdot p_{\pi }\right) \left(4 m_{\pi }^2-4 m_K^2\right) B_0\left(q,m_K,m_K\right)\\
&& +\frac{1}{9} \alpha _2 \left(8 m_K^2-8 m_{\pi }^2\right) \left(11 m_{\pi }^2-10 k_{\pi }\cdot p_{\pi }\right) B_0\left(q,m_{\pi },m_{\pi }\right)\\
&& +\frac{1}{9} \alpha _1 \left(-6 m_{\pi }^2-6 k_{\pi }\cdot p_K+6 k_{\pi }\cdot p_{\pi }+6 p_K\cdot p_{\pi }\right) \left(11 m_{\pi }^2-10 k_{\pi }\cdot p_{\pi }\right) B_0\left(q,m_{\pi },m_{\pi }\right)\\
&& +\frac{2}{9} \alpha _2 \left(4 m_K^2-4 m_{\pi }^2\right) B_0\left(q,m_{\eta },m_{\eta }\right) m_{\pi }^2+\frac{4}{9} \alpha _2 \left(8 m_{\pi }^2-8 m_K^2\right) B_1\left(q,m_{\pi },m_{\pi }\right)\\
&& +\frac{2}{9} \alpha _1 m_{\pi }^2 \left(3 m_{\pi }^2+3 k_{\pi }\cdot p_K-3 k_{\pi }\cdot p_{\pi }-3 p_K\cdot p_{\pi }\right) B_0\left(q,m_{\eta },m_{\eta }\right)\\
&& -\frac{2}{3} \alpha _1 \left(2 m_K^2+6 m_{\pi }^2+k_{\pi }\cdot p_K-6 k_{\pi }\cdot p_{\pi }-p_K\cdot p_{\pi }\right) B_1\left(q,m_K,m_K\right)\\
&& +\frac{4}{9} \alpha _1 \left(39 m_{\pi }^2+6 k_{\pi }\cdot p_K-36 k_{\pi }\cdot p_{\pi }-6 p_K\cdot p_{\pi }\right) B_1\left(q,m_{\pi },m_{\pi }\right)\\
&& -\frac{4}{3} \alpha _1 B_1\left(q,m_{\eta },m_{\eta }\right) m_{\pi }^2+\frac{4}{3} \alpha _1 B_2\left(q,m_K,m_K\right)-\frac{16}{3} \alpha _1 B_2\left(q,m_{\pi },m_{\pi }\right)\Big)
\end{eqnarray*}

\subsection{Diagram H1b}
Here, $ q = \frac{1}{2} ( p_K - k_\pi )$, 
\begin{eqnarray*}
\lefteqn{ \vev{ \pi^-( k_{\pi} ) | \co^{(8,1)} | K^0(p_K) \pi^-(p_{\pi} )}_{H1b} = \frac{1}{f^5 \pi^2} \Big( }&& \\ 
&& -3 \alpha _1 k_{\pi }\cdot p_K k_{\pi }\cdot B_0\left(q,m_K,m_{\pi }\right)+3 \alpha _1 k_{\pi }\cdot p_{\pi } k_{\pi }\cdot B_0\left(q,m_K,m_{\pi }\right)-\alpha _1 k_{\pi }\cdot p_K k_{\pi }\cdot B_0\left(q,m_{\eta },m_K\right)\\
&& +\alpha _1 k_{\pi }\cdot p_{\pi } k_{\pi }\cdot B_0\left(q,m_{\eta },m_K\right)-6 \alpha _1 k_{\pi }\cdot B_1\left(q,m_K,m_{\pi }\right)-2 \alpha _1 k_{\pi }\cdot B_1\left(q,m_{\eta },m_K\right)\\
&& -3 \alpha _1 k_{\pi }\cdot B_0\left(q,m_K,m_{\pi }\right) p_K\cdot p_{\pi }-\alpha _1 k_{\pi }\cdot B_0\left(q,m_{\eta },m_K\right) p_K\cdot p_{\pi }-3 \alpha _1 k_{\pi }\cdot p_K p_K\cdot B_0\left(q,m_K,m_{\pi }\right)\\
&& +3 \alpha _1 k_{\pi }\cdot p_{\pi } p_K\cdot B_0\left(q,m_K,m_{\pi }\right)-3 \alpha _1 p_K\cdot p_{\pi } p_K\cdot B_0\left(q,m_K,m_{\pi }\right)-\alpha _1 k_{\pi }\cdot p_K p_K\cdot B_0\left(q,m_{\eta },m_K\right)\\
&& +\alpha _1 k_{\pi }\cdot p_{\pi } p_K\cdot B_0\left(q,m_{\eta },m_K\right)-\alpha _1 p_K\cdot p_{\pi } p_K\cdot B_0\left(q,m_{\eta },m_K\right)-6 \alpha _1 p_K\cdot B_1\left(q,m_K,m_{\pi }\right)\\
&& -2 \alpha _1 p_K\cdot B_1\left(q,m_{\eta },m_K\right)-\frac{5}{3} \alpha _1 k_{\pi }\cdot p_K p_{\pi }\cdot B_0\left(q,m_K,m_{\pi }\right)+\frac{25}{3} \alpha _1 k_{\pi }\cdot p_K p_{\pi }\cdot B_0\left(q,m_{\eta },m_K\right)\\
&& +\frac{3}{2} \alpha _1 k_{\pi }\cdot B_0\left(q,m_K,m_{\pi }\right) m_K^2-\frac{2}{3} \alpha _1 p_{\pi }\cdot B_1\left(q,m_K,m_{\pi }\right)+\frac{10}{3} \alpha _1 p_{\pi }\cdot B_1\left(q,m_{\eta },m_K\right)\\
&& +\frac{8}{3} \alpha _2 k_{\pi }\cdot B_0\left(q,m_K,m_{\pi }\right) m_K^2+\frac{1}{2} \alpha _1 k_{\pi }\cdot B_0\left(q,m_{\eta },m_K\right) m_K^2+\frac{3}{2} \alpha _1 p_K\cdot B_0\left(q,m_K,m_{\pi }\right) m_K^2\\
&& +\frac{8}{3} \alpha _2 p_K\cdot B_0\left(q,m_K,m_{\pi }\right) m_K^2+\frac{1}{2} \alpha _1 p_K\cdot B_0\left(q,m_{\eta },m_K\right) m_K^2-\frac{5}{6} \alpha _1 p_{\pi }\cdot B_0\left(q,m_K,m_{\pi }\right) m_K^2\\
&& -\frac{85}{18} \alpha _1 p_{\pi }\cdot B_0\left(q,m_{\eta },m_K\right) m_K^2+\frac{3}{2} \alpha _1 k_{\pi }\cdot B_0\left(q,m_K,m_{\pi }\right) m_{\pi }^2-\frac{8}{3} \alpha _2 k_{\pi }\cdot B_0\left(q,m_K,m_{\pi }\right) m_{\pi }^2\\
&& +\frac{1}{2} \alpha _1 k_{\pi }\cdot B_0\left(q,m_{\eta },m_K\right) m_{\pi }^2+\frac{3}{2} \alpha _1 p_K\cdot B_0\left(q,m_K,m_{\pi }\right) m_{\pi }^2-\frac{8}{3} \alpha _2 p_K\cdot B_0\left(q,m_K,m_{\pi }\right) m_{\pi }^2\\
&& +\frac{1}{2} \alpha _1 p_K\cdot B_0\left(q,m_{\eta },m_K\right) m_{\pi }^2-\frac{5}{6} \alpha _1 p_{\pi }\cdot B_0\left(q,m_K,m_{\pi }\right) m_{\pi }^2-\frac{5}{18} \alpha _1 p_{\pi }\cdot B_0\left(q,m_{\eta },m_K\right) m_{\pi }^2\\
&& -2 \alpha _1 k_{\pi }\cdot B_0\left(q,m_K,m_{\pi }\right)\cdot p_{\pi }-10 \alpha _1 k_{\pi }\cdot B_0\left(q,m_{\eta },m_K\right)\cdot p_{\pi }-2 \alpha _1 p_{\pi }\cdot B_0\left(q,m_K,m_{\pi }\right)\cdot p_K\\
&& +\frac{1}{24} \alpha _1 \left(-m_K^2-m_{\pi }^2+2 k_{\pi }\cdot p_K-2 k_{\pi }\cdot p_{\pi }+2 p_K\cdot p_{\pi }\right) \left(50 k_{\pi }\cdot p_K-23 \left(m_K^2+m_{\pi }^2\right)\right) B_0\left(q,m_K,m_{\pi }\right)\\
&& +\frac{2}{9} \alpha _2 \left(m_K^2-m_{\pi }^2\right) \left(7 \left(m_K^2+m_{\pi }^2\right)-10 k_{\pi }\cdot p_K\right) B_0\left(q,m_K,m_{\pi }\right)\\
&& +\frac{1}{72} \alpha _1 \left(m_K^2+m_{\pi }^2-2 k_{\pi }\cdot p_K+2 k_{\pi }\cdot p_{\pi }-2 p_K\cdot p_{\pi }\right) \left(17 m_K^2+m_{\pi }^2-30 k_{\pi }\cdot p_K\right) B_0\left(q,m_{\eta },m_K\right)\\
&& +\frac{1}{3} \alpha _1 \left(-14 m_K^2-14 m_{\pi }^2+30 k_{\pi }\cdot p_K-5 k_{\pi }\cdot p_{\pi }+5 p_K\cdot p_{\pi }\right) B_1\left(q,m_K,m_{\pi }\right)\\
&& +\frac{1}{9} \alpha _1 \left(-10 m_K^2-2 m_{\pi }^2+18 k_{\pi }\cdot p_K-3 k_{\pi }\cdot p_{\pi }+3 p_K\cdot p_{\pi }\right) B_1\left(q,m_{\eta },m_K\right)\\
&& -\frac{8}{9} \alpha _2 \left(m_K^2-m_{\pi }^2\right) B_1\left(q,m_K,m_{\pi }\right)+\frac{10}{3} \alpha _1 B_2\left(q,m_K,m_{\pi }\right)+\frac{2}{3} \alpha _1 B_2\left(q,m_{\eta },m_K\right)\Big)
\end{eqnarray*}

\subsection{Diagram H2}
Here, $ q = -\frac{1}{2} ( p_K + p_\pi )$, 
\begin{eqnarray*}
\lefteqn{ \vev{ \pi^-( k_{\pi} ) | \co^{(8,1)} | K^0(p_K) \pi^-(p_{\pi} )}_{H2} = \frac{1}{f^5 \pi^2} \Big( }&& \\ 
&& \frac{4}{3} \alpha _1 k_{\pi }\cdot B_0\left(q,m_K,m_{\pi }\right) m_K^2-\frac{16}{3} \alpha _1 k_{\pi }\cdot B_1\left(q,m_K,m_{\pi }\right)+\frac{40}{3} \alpha _1 k_{\pi }\cdot B_0\left(q,m_K,m_{\pi }\right) p_K\cdot p_{\pi }\\
&& +\frac{4}{3} \alpha _1 k_{\pi }\cdot B_0\left(q,m_K,m_{\pi }\right) m_{\pi }^2-\frac{4}{9} \alpha _2 \left(m_K^2-m_{\pi }^2\right) \left(m_K^2+m_{\pi }^2+10 p_K\cdot p_{\pi }\right) B_0\left(q,m_K,m_{\pi }\right)\\
&& +\frac{16}{9} \alpha _2 \left(m_K^2-m_{\pi }^2\right) B_1\left(q,m_K,m_{\pi }\right)\Big)
\end{eqnarray*}

\subsection{Diagram H3}
\begin{eqnarray*}
\lefteqn{ \vev{ \pi^-( k_{\pi} ) | \co^{(8,1)} | K^0(p_K) \pi^-(p_{\pi} )}_{H3} = \frac{1}{f^5 \pi^2} \Big( }&& \\ 
&& \frac{64}{15} \left(-35 \alpha _1 k_{\pi }\cdot p_K+5 \alpha _1 p_K\cdot p_{\pi }+2 \left(25 \alpha _1 k_{\pi }\cdot p_{\pi }+7 \alpha _2 \left(m_K^2-m_{\pi }^2\right)\right)\right) A_0\left(m_K\right)\\
&& -\frac{32}{3} \left(-4 \alpha _2 m_K^2+4 \alpha _2 m_{\pi }^2+10 \alpha _1 k_{\pi }\cdot p_K-7 \alpha _1 k_{\pi }\cdot p_{\pi }+5 \alpha _1 p_K\cdot p_{\pi }\right) A_0\left(m_{\pi }\right)\\
&& +\frac{32}{45} \left(8 \alpha _2 m_K^2-8 \alpha _2 m_{\pi }^2-90 \alpha _1 k_{\pi }\cdot p_K+45 \alpha _1 k_{\pi }\cdot p_{\pi }+45 \alpha _1 p_K\cdot p_{\pi }\right) A_0\left(m_{\eta }\right)\\
&& -64 \alpha _1 A_1\left(m_K\right)+\frac{224}{3} \alpha _1 A_1\left(m_{\pi }\right)-\frac{32}{3} \alpha _1 A_1\left(m_{\eta }\right)\Big)
\end{eqnarray*}

\subsection{Diagram D12a}
Here, $ q = \frac{1}{2}( p_K + p_\pi ) $, 
\begin{eqnarray*}
\lefteqn{ \vev{ \pi^-( k_{\pi} ) | \co^{(8,1)} | K^0(p_K) \pi^-(p_{\pi} )}_{D12a} = \frac{\alpha_2(m_K^2 - m_\pi^2)}{4 f^5 \pi^2}  \frac{-1}{(p_k + p_\pi - k_\pi)^2 - m_K^2 } \Big( }&& \\ 
&& -\frac{1}{36} \left(m_K^2-3 m_{\pi }^2+4 k_{\pi }\cdot p_K+4 k_{\pi }\cdot p_{\pi }+6 p_K\cdot p_{\pi }\right) \left(m_K^2+m_{\pi }^2+10 p_K\cdot p_{\pi }\right) B_0\left(q,m_K,m_{\pi }\right)\\
&& +\frac{2}{9} \left(m_K^2-m_{\pi }^2+2 k_{\pi }\cdot p_K+2 k_{\pi }\cdot p_{\pi }+8 p_K\cdot p_{\pi }\right) B_1\left(q,m_K,m_{\pi }\right)\\
&& -\frac{4}{9} B_2\left(q,m_K,m_{\pi }\right)\Big)
\end{eqnarray*}

\subsection{Diagram D12b}
Here, $ q = \frac{1}{2}( k_\pi - p_\pi ) $, 
\begin{eqnarray*}
\lefteqn{ \vev{ \pi^-( k_{\pi} ) | \co^{(8,1)} | K^0(p_K) \pi^-(p_{\pi} )}_{D12b} = \frac{\alpha_2(m_K^2 - m_\pi^2)}{4 f^5 \pi^2}  \frac{-1}{(p_k + p_\pi - k_\pi)^2 - m_K^2 } \Big( }&& \\ 
&& -\frac{5}{3} p_K\cdot B_1\left(q,m_K,m_{\pi }\right)-p_K\cdot B_1\left(q,m_{\eta },m_K\right)-\frac{5}{3} p_{\pi }\cdot B_1\left(q,m_K,m_{\pi }\right)-p_{\pi }\cdot B_1\left(q,m_{\eta },m_K\right)\\
&& -\frac{1}{12} p_K\cdot B_0\left(q,m_K,m_{\pi }\right) \left(-23 m_K^2-13 m_{\pi }^2+40 k_{\pi }\cdot p_K-10 k_{\pi }\cdot p_{\pi }+10 p_K\cdot p_{\pi }\right)\\
&& +\frac{5}{6} k_{\pi }\cdot B_0\left(q,m_K,m_{\pi }\right) \left(-m_{\pi }^2+k_{\pi }\cdot p_K+k_{\pi }\cdot p_{\pi }-p_K\cdot p_{\pi }\right)\\
&& +\frac{1}{2} k_{\pi }\cdot B_0\left(q,m_{\eta },m_K\right) \left(-m_{\pi }^2+k_{\pi }\cdot p_K+k_{\pi }\cdot p_{\pi }-p_K\cdot p_{\pi }\right)\\
&& -\frac{1}{12} p_{\pi }\cdot B_0\left(q,m_{\eta },m_K\right) \left(-17 m_K^2-m_{\pi }^2+30 k_{\pi }\cdot p_K\right)\\
&& -\frac{1}{12} p_K\cdot B_0\left(q,m_{\eta },m_K\right) \left(-17 m_K^2+5 m_{\pi }^2+24 k_{\pi }\cdot p_K-6 k_{\pi }\cdot p_{\pi }+6 p_K\cdot p_{\pi }\right)\\
&& -\frac{1}{12} p_{\pi }\cdot B_0\left(q,m_K,m_{\pi }\right) \left(50 k_{\pi }\cdot p_K-23 \left(m_K^2+m_{\pi }^2\right)\right)\\
&& +\frac{3}{2} k_{\pi }\cdot B_0\left(q,m_K,m_{\pi }\right)\cdot k_{\pi }-3 k_{\pi }\cdot B_0\left(q,m_K,m_{\pi }\right)\cdot p_{\pi }+\frac{3}{2} k_{\pi }\cdot B_0\left(q,m_{\eta },m_K\right)\cdot k_{\pi }\\
&& -3 k_{\pi }\cdot B_0\left(q,m_{\eta },m_K\right)\cdot p_{\pi }-\frac{3}{2} p_K\cdot B_0\left(q,m_K,m_{\pi }\right)\cdot p_K-\frac{3}{2} p_K\cdot B_0\left(q,m_{\eta },m_K\right)\cdot p_K\\
&& -3 p_{\pi }\cdot B_0\left(q,m_K,m_{\pi }\right)\cdot p_K-3 p_{\pi }\cdot B_0\left(q,m_{\eta },m_K\right)\cdot p_K\\
&& -\frac{1}{32} \left(-3 m_K^2+m_{\pi }^2+6 k_{\pi }\cdot p_K-4 k_{\pi }\cdot p_{\pi }+4 p_K\cdot p_{\pi }\right) \left(10 k_{\pi }\cdot p_K-3 \left(m_K^2+m_{\pi }^2\right)\right) B_0\left(q,m_K,m_{\pi }\right)\\
&& -\frac{1}{144} \left(7 m_K^2+3 m_{\pi }^2-6 k_{\pi }\cdot p_K+4 k_{\pi }\cdot p_{\pi }-4 p_K\cdot p_{\pi }\right) \left(7 \left(m_K^2+m_{\pi }^2\right)-10 k_{\pi }\cdot p_K\right) B_0\left(q,m_K,m_{\pi }\right)\\
&& -\frac{1}{144} \left(7 m_K^2+3 m_{\pi }^2-6 k_{\pi }\cdot p_K+4 k_{\pi }\cdot p_{\pi }-4 p_K\cdot p_{\pi }\right) \left(7 \left(m_K^2+m_{\pi }^2\right)-10 k_{\pi }\cdot p_K\right) B_0\left(q,m_K,m_{\pi }\right)\\
&& -\frac{1}{864} \left(17 m_K^2-11 m_{\pi }^2-18 k_{\pi }\cdot p_K+12 k_{\pi }\cdot p_{\pi }-12 p_K\cdot p_{\pi }\right) \left(17 m_K^2+m_{\pi }^2-30 k_{\pi }\cdot p_K\right) B_0\left(q,m_{\eta },m_K\right)\\
&& +\frac{1}{4} \left(3 m_K^2+m_{\pi }^2-8 k_{\pi }\cdot p_K+2 k_{\pi }\cdot p_{\pi }-2 p_K\cdot p_{\pi }\right) B_1\left(q,m_K,m_{\pi }\right)\\
&& +\frac{1}{18} \left(7 m_K^2+5 m_{\pi }^2-8 k_{\pi }\cdot p_K+2 k_{\pi }\cdot p_{\pi }-2 p_K\cdot p_{\pi }\right) B_1\left(q,m_K,m_{\pi }\right)\\
&& -\frac{11}{18} B_2\left(q,m_K,m_{\pi }\right)-\frac{1}{6} B_2\left(q,m_{\eta },m_K\right)\Big)
\end{eqnarray*}

\subsection{Diagram D12c}
Here, $ q = \frac{1}{2}( p_K - k_\pi  ) $, 
\begin{eqnarray*}
\lefteqn{ \vev{ \pi^-( k_{\pi} ) | \co^{(8,1)} | K^0(p_K) \pi^-(p_{\pi} )}_{D12c} = \frac{\alpha_2(m_K^2 - m_\pi^2)}{4 f^5 \pi^2}  \frac{-1}{(p_k + p_\pi - k_\pi)^2 - m_K^2 } \Big( }&& \\ 
&& k_{\pi }\cdot B_0\left(q,m_K,m_K\right)\cdot k_{\pi }-2 k_{\pi }\cdot B_0\left(q,m_K,m_K\right)\cdot p_K+2 k_{\pi }\cdot B_0\left(q,m_{\pi },m_{\pi }\right)\cdot k_{\pi }\\
&& -4 k_{\pi }\cdot B_0\left(q,m_{\pi },m_{\pi }\right)\cdot p_K-2 p_{\pi }\cdot B_0\left(q,m_K,m_K\right)\cdot p_K-p_{\pi }\cdot B_0\left(q,m_K,m_K\right)\cdot p_{\pi }\\
&& -4 p_{\pi }\cdot B_0\left(q,m_{\pi },m_{\pi }\right)\cdot p_K-2 p_{\pi }\cdot B_0\left(q,m_{\pi },m_{\pi }\right)\cdot p_{\pi }\\
&& +\frac{1}{12} \left(-2 m_K^2-5 m_{\pi }^2+5 k_{\pi }\cdot p_{\pi }\right) \left(2 m_K^2+3 m_{\pi }^2+2 k_{\pi }\cdot p_K-3 k_{\pi }\cdot p_{\pi }-2 p_K\cdot p_{\pi }\right) B_0\left(q,m_K,m_K\right)\\
&& +\frac{1}{18} \left(5 k_{\pi }\cdot p_{\pi }-7 m_{\pi }^2\right) \left(5 m_{\pi }^2+2 k_{\pi }\cdot p_K-3 k_{\pi }\cdot p_{\pi }-2 p_K\cdot p_{\pi }\right) B_0\left(q,m_{\pi },m_{\pi }\right)\\
&& +\frac{1}{18} \left(5 k_{\pi }\cdot p_{\pi }-4 m_{\pi }^2\right) \left(5 m_{\pi }^2+2 k_{\pi }\cdot p_K-3 k_{\pi }\cdot p_{\pi }-2 p_K\cdot p_{\pi }\right) B_0\left(q,m_{\pi },m_{\pi }\right)\\
&& -\frac{1}{18} m_{\pi }^2 \left(7 m_{\pi }^2+6 k_{\pi }\cdot p_K-9 k_{\pi }\cdot p_{\pi }-6 p_K\cdot p_{\pi }\right) B_0\left(q,m_{\eta },m_{\eta }\right)\\
&& +\frac{1}{3} \left(2 m_K^2+4 m_{\pi }^2+k_{\pi }\cdot p_K-4 k_{\pi }\cdot p_{\pi }-p_K\cdot p_{\pi }\right) B_1\left(q,m_K,m_K\right)\\
&& +\frac{2}{9} \left(6 m_{\pi }^2+k_{\pi }\cdot p_K-4 k_{\pi }\cdot p_{\pi }-p_K\cdot p_{\pi }\right) B_1\left(q,m_{\pi },m_{\pi }\right)\\
&& +\frac{1}{9} \left(9 m_{\pi }^2+2 k_{\pi }\cdot p_K-8 k_{\pi }\cdot p_{\pi }-2 p_K\cdot p_{\pi }\right) B_1\left(q,m_{\pi },m_{\pi }\right)\\
&& +\frac{1}{3} B_1\left(q,m_{\eta },m_{\eta }\right) m_{\pi }^2-\frac{1}{3} B_2\left(q,m_K,m_K\right)-\frac{4}{9} B_2\left(q,m_{\pi },m_{\pi }\right)\Big)
\end{eqnarray*}

\subsection{Diagram D3}
\begin{eqnarray*}
\lefteqn{ \vev{ \pi^-( k_{\pi} ) | \co^{(8,1)} | K^0(p_K) \pi^-(p_{\pi} )}_{D3} = \frac{\alpha_2(m_K^2 - m_\pi^2)}{4 f^5 \pi^2}  \frac{-1}{(p_k + p_\pi - k_\pi)^2 - m_K^2 } \Big( }&& \\ 
&& \frac{32}{45} \left(-18 k_{\pi }\cdot k_{\pi }+13 k_{\pi }\cdot p_K+18 k_{\pi }\cdot p_{\pi }+17 p_K\cdot p_K+47 p_K\cdot p_{\pi }+12 p_{\pi }\cdot p_{\pi }\right) A_0\left(m_K\right)\\
&& +\frac{16}{9} \left(-8 k_{\pi }\cdot k_{\pi }+13 k_{\pi }\cdot p_K+8 k_{\pi }\cdot p_{\pi }+2 p_K\cdot p_K+17 p_K\cdot p_{\pi }+7 p_{\pi }\cdot p_{\pi }\right) A_0\left(m_{\pi }\right)\\
&& +\frac{16}{15} \left(-4 k_{\pi }\cdot k_{\pi }-k_{\pi }\cdot p_K+4 k_{\pi }\cdot p_{\pi }+6 p_K\cdot p_K+11 p_K\cdot p_{\pi }+p_{\pi }\cdot p_{\pi }\right) A_0\left(m_{\eta }\right)\\
&& +\frac{544}{45} A_1\left(m_K\right)+\frac{112}{9} A_1\left(m_{\pi }\right)+\frac{16}{15} A_1\left(m_{\eta }\right)\Big)
\end{eqnarray*}

\subsection{Diagram F1}
\begin{eqnarray*}
\lefteqn{ \vev{ \pi^+( k_{\pi} ) | \co^{(8,1)} | K^+(p_K)  )}_{F3} = \frac{i}{f^4 \pi^2} \Big( }&& \\ 
&& -\frac{64}{3} \left(7 \alpha _1 k_{\pi }\cdot p_K-4 \alpha _2 m_K^2\right) A_0\left(m_K\right)-\frac{160}{3} \left(2 \alpha _1 k_{\pi }\cdot p_K-\alpha _2 m_K^2\right) A_0\left(m_{\pi }\right)\\
&& -\frac{32}{3} \left(6 \alpha _1 k_{\pi }\cdot p_K-\alpha _2 m_K^2\right) A_0\left(m_{\eta }\right)-64 \alpha _1 A_1\left(m_K\right)+\frac{64}{3} \alpha _1 A_1\left(m_{\pi }\right)\Big)
\end{eqnarray*}

\subsection{Diagram F2}
Here, $ q = \frac{1}{2}( k_\pi-p_\pi  ) $, 
\begin{eqnarray*}
\lefteqn{ \vev{ \pi^+( k_{\pi} ) | \co^{(8,1)} | K^+(p_K)  )}_{F2} = \frac{i}{f^4 \pi^2} \Big( }&& \\ 
&& 6 \alpha _1 k_{\pi }\cdot B_1\left(q,m_K,m_{\pi }\right)-2 \alpha _1 k_{\pi }\cdot B_1\left(q,m_K,m_{\eta }\right)+6 \alpha _1 p_K\cdot B_1\left(q,m_K,m_{\pi }\right)\\
&& -\frac{3}{2} \alpha _1 k_{\pi }\cdot B_0\left(q,m_K,m_{\pi }\right) m_{\pi }^2+\frac{1}{2} \alpha _1 k_{\pi }\cdot B_0\left(q,m_K,m_{\eta }\right) m_{\pi }^2-2 \alpha _1 p_K\cdot B_1\left(q,m_K,m_{\eta }\right)\\
&& +\frac{1}{2} \alpha _1 p_K\cdot B_0\left(q,m_K,m_{\eta }\right) m_{\pi }^2-\frac{3}{2} \alpha _1 p_K\cdot B_0\left(q,m_K,m_{\pi }\right) m_{\pi }^2\\
&& +\frac{1}{24} \alpha _1 m_{\pi }^2 \left(8 m_K^2+13 m_{\pi }^2-20 k_{\pi }\cdot p_K+10 k_{\pi }\cdot p_{\pi }-10 p_K\cdot p_{\pi }\right) B_0\left(q,m_K,m_{\pi }\right)\\
&& -\frac{1}{72} \alpha _1 m_{\pi }^2 \left(-8 m_K^2+5 m_{\pi }^2+12 k_{\pi }\cdot p_K-6 k_{\pi }\cdot p_{\pi }+6 p_K\cdot p_{\pi }\right) B_0\left(q,m_K,m_{\eta }\right)\\
&& +\frac{1}{3} \alpha _1 \left(-4 m_K^2-9 m_{\pi }^2+10 k_{\pi }\cdot p_K-5 k_{\pi }\cdot p_{\pi }+5 p_K\cdot p_{\pi }\right) B_1\left(q,m_K,m_{\pi }\right)\\
&& +\frac{1}{9} \alpha _1 \left(-4 m_K^2+m_{\pi }^2+6 k_{\pi }\cdot p_K-3 k_{\pi }\cdot p_{\pi }+3 p_K\cdot p_{\pi }\right) B_1\left(q,m_K,m_{\eta }\right)\\
&& +\frac{10}{3} \alpha _1 B_2\left(q,m_K,m_{\pi }\right)+\frac{2}{3} \alpha _1 B_2\left(q,m_K,m_{\eta }\right)\Big)
\end{eqnarray*}

\subsection{Diagram A3}
\begin{eqnarray*}
\lefteqn{ \vev{ \pi^-( k_{\pi} ) | \co^{(8,1)} | K^0(p_K) \pi^-(p_{\pi} )}_{A3} = \frac{1}{f^5 \pi^2} \Big( }&& \\ 
&& -\frac{128 \alpha _2 \left(-2 m_{\pi }^2+2 k_{\pi }\cdot p_K+2 k_{\pi }\cdot p_{\pi }+4 p_K\cdot p_{\pi }\right) \left(m_K^2-m_{\pi }^2\right) A_0\left(m_K\right)}{3 \left(2 m_{\pi }^2-2 k_{\pi }\cdot p_K-2 k_{\pi }\cdot p_{\pi }+2 p_K\cdot p_{\pi }\right)}\\
&& -\frac{128 \alpha _2 \left(-2 m_{\pi }^2+2 k_{\pi }\cdot p_K+2 k_{\pi }\cdot p_{\pi }+4 p_K\cdot p_{\pi }\right) \left(m_K^2-m_{\pi }^2\right) A_0\left(m_K\right)}{3 \left(2 m_{\pi }^2-2 k_{\pi }\cdot p_K-2 k_{\pi }\cdot p_{\pi }+2 p_K\cdot p_{\pi }\right)}\\
&& -\frac{64 \alpha _2 \left(-2 m_{\pi }^2+2 k_{\pi }\cdot p_K+2 k_{\pi }\cdot p_{\pi }+4 p_K\cdot p_{\pi }\right) \left(m_K^2-m_{\pi }^2\right) A_0\left(m_{\pi }\right)}{3 \left(2 m_{\pi }^2-2 k_{\pi }\cdot p_K-2 k_{\pi }\cdot p_{\pi }+2 p_K\cdot p_{\pi }\right)}\\
&& -\frac{64 \alpha _2 \left(-2 m_{\pi }^2+2 k_{\pi }\cdot p_K+2 k_{\pi }\cdot p_{\pi }+4 p_K\cdot p_{\pi }\right) \left(m_K^2-m_{\pi }^2\right) A_0\left(m_{\eta }\right)}{9 \left(2 m_{\pi }^2-2 k_{\pi }\cdot p_K-2 k_{\pi }\cdot p_{\pi }+2 p_K\cdot p_{\pi }\right)}\\
&& +\frac{64 \alpha _1 \left(-2 m_{\pi }^2+2 k_{\pi }\cdot p_K+2 k_{\pi }\cdot p_{\pi }+4 p_K\cdot p_{\pi }\right) A_1\left(m_K\right)}{3 \left(2 m_{\pi }^2-2 k_{\pi }\cdot p_K-2 k_{\pi }\cdot p_{\pi }+2 p_K\cdot p_{\pi }\right)}\\
&& -\frac{32 \alpha _1 \left(-2 m_{\pi }^2+2 k_{\pi }\cdot p_K+2 k_{\pi }\cdot p_{\pi }+4 p_K\cdot p_{\pi }\right) A_1\left(m_{\pi }\right)}{2 m_{\pi }^2-2 k_{\pi }\cdot p_K-2 k_{\pi }\cdot p_{\pi }+2 p_K\cdot p_{\pi }}\\
&& +\frac{32 \alpha _1 \left(-2 m_{\pi }^2+2 k_{\pi }\cdot p_K+2 k_{\pi }\cdot p_{\pi }+4 p_K\cdot p_{\pi }\right) A_1\left(m_{\eta }\right)}{3 \left(2 m_{\pi }^2-2 k_{\pi }\cdot p_K-2 k_{\pi }\cdot p_{\pi }+2 p_K\cdot p_{\pi }\right)}\Big)
\end{eqnarray*}

\section{Finite Volume Effects}
\label{sec:apnd_FVE}

The following argument follows Ref.\cite{Lin:2001fi}, where more detailed discussion can be found. The relevant correlation function is
\begin{eqnarray}
\lefteqn{
\cC(t) \equiv
\int_V d^x e^{\bP \cdot \bx} \vev{\pi(\bk_\pi) |   H_W(0)  \cO_{K\pi}(t,x) | 0  } 
} &&
 \\
&&
\approx V \sum_n 
\vev{\pi(\bk_\pi) | H_W(0)  |  K\pi,n,\bP} e^{-E_n t}
\vev{ K\pi,n,\bP| \cO_{K\pi}(0) | 0 }
\nonumber
\end{eqnarray}
where contributions from excited state {\it e.g.} four particle states are ignored since we are interested in asymptotic limit in which $ t \to \infty$.

If the volume is sufficiently large, the summation can be approximated
 with integration :
\begin{eqnarray}
\lefteqn{
\cC(t)=
} &&
\label{eq:ME_V}
 \\
&&
V \int_0^\infty dE ~
\rho_V(E) e^{-Et}
\vev{\pi(\bk_\pi) | H_W(0)  |  K\pi,(E,\bP)} 
\vev{ K\pi,(E,\bP)| \cO_{K\pi}(0) | 0 },
\nonumber
\end{eqnarray}
\noindent
where the interpretation of $\rho_V(E)$ is given in Ref.\cite{Lin:2001fi}. Roughly, $\rho_V$ can be understood as density of states. 

Meanwhile, the large volume allows us to rewrite the correlation function, $\cC(t)$, in terms of infinite volume asymptotic states:
\begin{eqnarray}
\lefteqn{
\cC(t)= 
} &&
\label{eq:ME_C}
\\
&&
\left(\frac{1}{2\pi}\right)^6 \int \frac{ dp_1^3}{2E_1 } \frac{ dp_2^3}{2E_2}
\vev{ \pi(\bk_\pi)| H_W(0) | K(\bp_1) \pi(\bp_2) }
\vev{  K(\bp_1) \pi(\bp_2)| \cO_{K\pi} | 0 }.
\nonumber 
\end{eqnarray}
We perform a small manipulation on the above equation using
\begin{eqnarray*}
\vev{ K(\bp_K) \pi(\bp_\pi)| \cO_{K\pi}(t,\bP) |  0}
&=&
\int d^3 x e^{i \bP \cdot \bx }
\vev{ K(\bp_K) \pi(\bp_\pi) | \cO_{K\pi}(t,\bx) | 0} \\
&=&
\int d^3 x e^{i \bP \cdot \bx }
  e^{-i  (\bp_K + \bp_\pi ) \cdot \bx} e^{- E t }
\vev{ K(\bp_K) \pi(\bp_\pi)| \cO_{K\pi}(0) | 0 } \\
&=&
(2\pi)^3 \delta( \bP - \bp_K - \bp_\pi ) 
\vev{ K(\bp_K) \pi(\bp_\pi) | \cO_{K\pi}(0) | 0} e^{- E t }
\end{eqnarray*}
where
\begin{equation}
E=E_K(\bp_K) + E_\pi(\bp_\pi),  ~~~~  E_i(\bp) = \sqrt{m_i^2 + \bp^2}.
\end{equation}
Then the correlation function can be written in terms of 4-momentum integral:
\begin{eqnarray*}
 \lefteqn{\vev{ \pi(\bk_\pi)| H_W(0) \cO_{K\pi}(t,\bP)   | 0 }=}
&&\\
&&
(2\pi)^3\int dE \int \frac{ dp_K^4}{(2\pi)^4} \frac{ dp_\pi^4}{(2\pi)^4}
 \delta^+(p_K^2 - m_K^2) \delta^+(p_\pi^2 - m_\pi^2 )
 \delta^4(P - p_K - p_\pi ) e^{- (E_K + E_\pi) t }\\
&& \hspace{2cm} \vev{ \pi(\bk_\pi) | H_W(0) | K(p_K) \pi(p_\pi)}
\vev{ K(p_K) \pi(p_\pi)| \cO_{K\pi}(0) | 0 }
\end{eqnarray*}
where
\begin{equation}
p_i=(E_i,\bp_i).
\end{equation}
Since the integral is in covariant form,
one can easily change the integration variables with Lorentz transformed ones  
 which brings the two particle state into CM frame:
\begin{eqnarray*}
\lefteqn{\vev{ \pi(\bk_\pi)|  H_W(0) \cO_{K\pi}(t,\bP)  |0 }=}
&&\\
&&
(2\pi)^3\int dE \int \frac{ dq_K^4}{(2\pi)^4} \frac{ dq_\pi^4}{(2\pi)^4}
 \delta^+(q_K^2 - m_K^2) \delta^+(q_\pi^2 - m_\pi^2 )
 \delta(E^* - E_K^* - E_\pi^*) \delta^3( \bq_K + \bq_\pi ) e^{- E t}\\
&& \hspace{2cm} \vev{ \pi(\bk_\pi^*) | H_W(0) | K(q_K) \pi(q_\pi)}
\vev{ K(q_K) \pi(q_\pi)| \cO_{K\pi}(0) |  0}.
\end{eqnarray*}
Note that the amplitudes are treated as scalars.
Using the $\delta$-function, one can do some integrations,
\begin{eqnarray*}
\lefteqn{\vev{ \pi(\bk_\pi)|  H_W(0) \cO_{K\pi}(t,\bP)  | 0 }=}
&&\\
&&
(2\pi)^{-3} \int dE \,\, e^{-Et} \int \frac{ dq^3}{2E_K(\bq) \, 2E_\pi(\bq)} 
 \delta(E^* - E_K(\bq) - E_\pi(\bq))  \\
&& \hspace{2cm} \vev{ \pi(\bk_\pi^*)| H_W(0) | K(\bq) \pi(-\bq) }
\vev{ K(\bq) \pi(-\bq) | \cO_{K\pi}(0) | 0},
\end{eqnarray*}
\noindent where $E^*_K$ and $E^*_\pi$ are fixed respectively by
\begin{equation}
E_K(\bq) = \sqrt{m_K^2 + \bq^2}, ~~~~~~~   E_\pi(\bq) = \sqrt{m_\pi^2 + \bq^2}.
\end{equation}

In order to proceed, we have to assume that 
the creation operator $\cO_{K\pi}$ is chosen so that
$\vev{ K(\bq) \pi(-\bq)| \cO_{K\pi}(0) | 0}$ does not have angular dependence.
 In fact, the contribution originating from the particular form of $O_{K\pi}$ will be canceled in the end.
 The role of $O_{K\pi}$ is restricted to keeping the initial states in S-wave.
 With this assumption, a further simplification is possible :
\begin{eqnarray}
\lefteqn{\vev{ \pi(\bk_\pi) |  H_W(0) \cO_{K\pi}(t,\bP) | 0}=}
&& \nonumber \\
&&
(2\pi)^{-3}\int dE \,\, e^{-Et}
\frac{1}{q^*}  \frac{E_K(q^*)  E_\pi(q^*)}{ E_K(q^*) + E_\pi(q^*) } 
  \vev{  K(\bq^*) \pi(-\bq^*) | \cO_{K\pi}(0) | 0 }
\nonumber
\\
&& \hspace{2cm}\int d\Omega \vev{  \pi(\bk_\pi^*)| H_W(0) |K(\bq) \pi(-\bq) },
\label{eq:ME_C2}
\end{eqnarray}
where $q^*$ is defined as
\begin{equation} 
E^* = \sqrt{ m_K^2 + q^{*2}} + \sqrt{ m_\pi^2 + q^{*2}}.
\end{equation}
Since  $\vev{  K(\bq^*) \pi(-\bq^*) | \cO_{K\pi}(0) | 0 }$ does not have any angular dependence, we write this as $\vev{  K \pi , E^* | \cO_{K\pi}(0) | 0 }$. 
Moreover, $\int d\Omega \vev{  \pi(\bk_\pi^*)| H_W(0) |K(\bq) \pi(-\bq) }$ is also a function of $E^*$ because of angular averaging so we use a notation $d\Omega_{E^*}$.

By comparing Eq.(\ref{eq:ME_V}) and Eq.(\ref{eq:ME_C2}), one can deduce
\begin{eqnarray}
\lefteqn{\rho_V(E) \vev{ \pi(\bk_\pi)| H_W(0)  | K\pi,(E,\bP) }
\vev{ K\pi,(E,\bP) | \cO_{K\pi} | 0}
=} \label{eq:FV_REL}
&& \\
&&\frac{1}{4 (2\pi)^3 } \left(\frac{q}{E^*} \right) 
 \vev{ K \pi , E^* | \cO_{K\pi}(0) | 0}
\int d\Omega_{E^*} \vev{  \pi(k_\pi^*)| H_W(0) |K(\bq) \pi(-\bq) }.
\nonumber
\end{eqnarray}

However, Eq.(\ref{eq:FV_REL}) is not satisfactory because of the appearance of an unknown factor $ \vev{ K \pi , E^* | \cO_{K\pi}(0) | 0}$.
In order to eliminate this factor, we consider a correlation function,
\begin{equation} 
\vev{ \cO_{K\pi}(t,\bP) \cO^\dagger_{K\pi}(0,\bP) }.
\end{equation}
One can easily imagine that by the similar argument, the following equation can be obtained,
\begin{eqnarray}
\rho_V(E) |\vev{  K\pi,(E,\bP) | \cO_{K\pi} |0}|^2
=
\frac{\pi}{ (2\pi)^3 }
\left(\frac{q^*}{E^*} \right) 
 |\vev{ K \pi , E^*  | \cO_{K\pi}(0) |0}|^2. 
\label{eq:FV_KPI}
\end{eqnarray}
By dividing Eq.(\ref{eq:FV_REL}) by the square root of Eq.(\ref{eq:FV_KPI}), the relation between matrix elements can be deduced,
\begin{eqnarray*}
\lefteqn{ \Big|\vev{\pi(\bk_\pi) | H_W(0)  | K\pi,(E,\bP) }\Big| = }&& \\
&& \frac{1}{4\pi} \frac{1}{\sqrt{\rho_V}}
\sqrt{ \frac{\pi}{(2\pi)^3 } \left(\frac{q^*}{E^*} \right)} 
\Big|\int d\Omega_{E^*}
 \vev{  \pi(\bk_\pi^*)| H_W(0) |K(\bq) \pi(-\bq) }\Big|.
\end{eqnarray*}

\section{LECs for $\cO^{\Delta I=1/2}_{(27,1)}$ and $\cO^{\Delta I=1/2}_{(8,8)}$}
\label{apnd:C}

It has been argued that LECs for operators with (8,8) $\Delta I=1/2$ 
 can determined without $K\to\pi\pi_{I=0}$ in Ref.\cite{Laiho:2002jq}.
In this section, we will show that LECs for operators with (27,1) $\Delta I=1/2$ as well 
 can be determined without calculating $\Delta I=1/2$, $K\to\pi\pi_{I=0}$ matrix elements.
Its analytic term at physical kinematics is given by
\begin{eqnarray} 
\lefteqn{-\frac{4 i (m_K^2-m_{\pi }^2)} {f^3}
\Big[-\alpha_{27} + \left(d_4+d_5-9 d_6+4 d_7\right) m_K^2 } & & \nonumber \\
& & +2 \left(-6 d_1-2 d_2+2 d_4+6 d_6+d_7-2
   d_{20}+8 d_{24}\right) m_{\pi }^2\Big].
\label{eq:phys_Kpipi}
\end{eqnarray}

The $K\to\pi$ transition amplitude with $(27,1)$ $\Delta I=3/2$,  are : 
\begin{eqnarray} 
\lefteqn{\frac{1}{f^2} \Big[ -4 \alpha_{27}(p_K \cdot k_\pi) + 64 d_{24} (p_K \cdot  k_\pi)^2
+8 \left(d_4+d_7-d_{20}\right) m_{\pi }^2 (p_K \cdot k_\pi)}& & \nonumber \\
& &+m_K^2 \left(8
   \left(d_4+2 d_7-d_{20}\right) (p_K \cdot k_\pi)-16 d_2 m_{\pi }^2\right) \Big].
\end{eqnarray}
From the above, one can determine $\alpha_{27},d_2,d_7,d_{20}-d_4,d_{24}$ by varying masses and momenta.

The $\chi$PT formula for $K\to\pi\pi_{I=2} $ transition amplitude with unphysical kinematics corresponding to the initial kaon and the final pions are at rest, is given by
\begin{eqnarray} 
\lefteqn{-\frac{8 i m_{\pi }}{f^3}\Big[ -\alpha_{27} (m_K+m_\pi)/2 + \left(d_4+d_5+4 d_7-d_{20}\right) m_K^3}& & \nonumber
\\
& &+2 \left(d_{20}-d_2\right) m_{\pi } m_K^2+\left(3 d_4+d_5+2
   d_7-3 d_{20}\right) m_{\pi }^2 m_K+2 d_2 m_{\pi }^3\Big].
\nonumber \\
\end{eqnarray}
By inspecting the above formula, we can see that $d_5 - d_4$ can be determined.
Since the $\chi$PT formula for the $K\to\vac$ transition amplitude is 
 \begin{equation} 
\frac{48 i d_1 \left(m_K^2-m_{\pi }^2\right)^2}{f},
\end{equation}
\noindent
$d_1$ can be fixed from this.
Finally, $d_6$ can be determined by using $\Delta I=1/2$ $K\to\pi$ transition amplitude whose $\chi$PT formula is
\begin{eqnarray} 
\lefteqn{\frac{1}{f^2} \Big[ -4 \alpha_{27}(p_K \cdot k_\pi) +  \left(8 \left(d_4-3 d_6+2 d_7-d_{20}\right) (p_K \cdot k_\pi)-16 \left(3 d_1+d_2\right)
   m_{\pi }^2\right) m_K^2 } & &  \nonumber \\
& &
+48 d_1 m_K^4
+64 d_{24} (p_K \cdot k_\pi)^2+8 \left(d_4+3 d_6+d_7-d_{20}\right)
   (p_K \cdot k_\pi) m_{\pi }^2 \Big]. \nonumber \\
\end{eqnarray}

Since $\alpha_{27},d_1,d_2,d_5-d_4,d_6,d_7,d_{20}-d_4,d_{24}$ are 
 determined, one can reconstruct the physical $K\to\pi\pi$ transition amplitude with $(27,1)$ $\Delta I=1/2$ using Eq.(\ref{eq:phys_Kpipi}).


%
%

\end{document}